# The Extended Evolutionary Synthesis Facilitates Evolutionary Models of Culture Change.

Cameron M. Smith[1], Liane Gabora[2] and William Gardner-O'Kearney[1]

1       Department of Anthropology, Portland State University
2       Department of Psychology, University of British Columbia

## Abstract

The Extended Evolutionary Synthesis (EES) is beginning to fulfill the whole promise of Darwinian insight through its extension of evolutionary understanding from the biological domain to include cultural information evolution. Several decades of important foundation-laying work took a social Darwinist approach and exhibited and ecologically-deterministic elements. This is not the case with more recent developments to the evolutionary study of culture, which emphasize non-Darwinian processes such as self-organization, potentiality, and epigenetic change.

## 1.    Introduction

In recent years, *Nature* editors were compelled to declare evolution a fact (2008), and a 2010 *Proceedings of the Royal Society [B]* symposium *Culture Evolves* declared culture to be an evolutionary process. The adoption of evolutionary approaches to culture has in part been spurred by the recognition that the currently-developing Extended Evolutionary Synthesis (Danchin et al., 2011; Koonin, 2009; Love, 2010; Pigliucci and Muller, 2010) provides a more textured appreciation for the multiple modes of evolution, including cultural evolution (Smith and Ruppell, 2011). In this paper, we outline an evolutionary approach to culture that is free of the social-Darwinism and ecologically-deterministic elements that characterized many earlier approaches. We use the term *evolution* to refer to a process that is cumulative, adaptive, and open-ended, and culture is all of these; i.e., culture evolves (Boyd & Richerson, 1985; Gabora, 1995; Whiten et al., 2011). We note that while some authors use the term 'culture change', we believe that is misleading, for 'change' need not be cumulative, adaptive, and open-ended.[1]

Critiques of evolutionary models of culture have a long history in the Americanist anthropological tradition (Carneiro, 2003; Mace, 2014; Perry and Mace, 2010), and today there remains question about the appropriateness of the 'analogy' between cultural and biological evolution (Claidière and André, 2011; Gabora, 2013). Cultural evolution uses different information channels, with different properties.

Note that while some view the central criterion of evolution to be replication with variation and selection (e.g., Hull et al., 2001), this is but one form of evolution. Evolution can also occur through *communal exchange and self-organization* (Gabora, 2013; Vetsigian, 2006) and through *context-driven actualization of potential* (Gabora, 2005, 2006) (for specific and general discussions of this topic see Kopps et al., 2015 and Gabora and Aerts, 2002, respectively; see also *Appendix 1*.) This approach is sometimes referred to as Self-Other Reorganization, because it involves both interactions *within* self-organizing structures, and interactions *between* them. We emphasize that for a process to be *evolutionary* (whether it be Darwinian evolution, or not), change must occur *on the basis of a fitness function,* or an environment that confers constraints and affordances. If not, i.e., if change is random, it is not due to evolution but to processes such as drift (i.e., variation in the relative frequency of different genotypes in a small population, owing to the chance disappearance of particular genes as individuals die or do not reproduce). Cultural evolution is fueled by the generation of, and reflection on, creative ideas, which may exist not in the form of a collection of explicitly actualized variants as is required for biological evolution, but in a state of potentiality[2] (Gabora, 2017). If an idea in a state of potentiality is considered with respect to one context it evolves one way, whereas if considered with respect to another context it evolves another way; there are no variants that get actualized and selected amongst. The mathematical description of evolution through variation and selection is very different from that of evolution through actualization of potentiality, which can be mathematically described drawing on the formalisms of superposition and interference (this is explored in the following literature: Aerts et al., 2016; Gabora and Aerts, 2005; Gabora and Carbert, 2015; Gabora and Saab, 2011).

---

[1] An asteroid *changes* as it moves across the universe – little particles might chip off for example, and it changes its spatial coordinates – but it does not *evolve*. To use the word change is to imply that culture is nothing more spectacular than what the asteroid undergoes.

[2] For example, let's say the cultural output in question is an idea for a screenplay. If you were to think about it from your mother's perspective it might come out one way, while if you were to think about it from your best friends perspective, it might come out another way. The different ways it could have manifested never actually exist as simultaneously actualized movies or scripts in a 'generation' of variant scripts, with the fittest being 'selected' and the least fit discarded. It simply exists in a state of potentiality that could manifest different ways, and over time it takes shape in one of these specific ways.

The principal differences between biological and cultural information (e.g. see Richerson et al., 2010) are addressed by the EES. For example, cultural information *has the potential* to evolve faster than biological information (e.g., Reynolds, 1994; Gabora, 1997), proposed by some to result in genetic evolution lagging behind cultural evolution in the face of selective pressure change. An example of this can be found in dietary changes that have arisen culturally since the Neolithic, for which the human genome has not yet fully responded (Arnold, 2014), with phenotypic plasticity maintaining fitness in the interim (Perreault, 2012).

Another major difference between cultural and biological evolution is that culture (extrasomatic information) may be transmitted horizontally, among members of a given generation, and in so being has long been called fundamentally non-evolutionary in its processes. However, *horizontal gene transfer* (discussed further below) is prevalent in the world of the asexually-reproducing species, and has been since lifebegan billions of years ago (Bock, 2010; Dunning Hotopp, 2011; McDaniel et al., 2010; Syvanen, 2012). Thus, with respect to the *Inheritance of Acquired Characteristics* sense, Lamarck was broadly correct about a fundamental evolutionary mechanism for most life (which is microbial) and for all of the history of life on Earth--and in the case of cultural information, horizontal transmission of information has been important since at least the time of the most recent evolutionary transition (*sensu* Szathmary and Maynard-Smith) which included the evolution of complex, learned and shared extrasomatic guides to behavior, also known as 'culture'.

Finally, it has been convincingly argued that ecologically-deterministic models of cultural selection that do not account for the variability of human behavior are unrealistically crude, reducing primate individuals to Optimal Foragers slaved to fitness calculations (e.g., Laland, 2015). However, EES-influenced workers are responding; Gabora (1999, 2013) has proposed an evolutionary (in the above sense) albeit non-Darwinian model of culture that highlights individual agency in an evolutionary framework.

Below, we identify specific reasons for building an evolutionary theory of culture, and show how certain aspects of the EES are contributing to this aim.

## 2.    What an Evolutionary Model of Culture Can Explain

Before the advent of the EES, Durham (1991) listed three reasons for developing a 'sequential transformation theory of cultural change': (1) to give a realistic time dimension to living cultures, (2) to use this dimension to understand "…the historical processes through which people have composed, edited and revised the [symbols] that give meaning and direction to their lives", and (3), to "account for trends in the historical emergence and divergence of ideational systems" (pp. 31-32). What specifically could such a theory of cultural evolution explain? The term 'culture' has been much-debated in anthropology (Kronfeldner, 2010; Mesoudi et al., 2006; Rohner, 1984), but for our purposes it srefers to nongenetic information, used in the shaping of behavior, transmitted among (and down generations of) members of groups; that is, learned, shared guides to behavior contrasting to instinctually-directed behavior. Cultures differ, of course, but so do varieties of biological organisms which have not been impossible to analyze and understand with evolutionary tools. For example, Love (2010) identified a number of 'stable elements' or recurring themes explored in multiple widely-used evolutionary biology texts (Table 1, Column 1): these are what evolution is used to explain in biology. Similarly, a review of several modern, widely-used cultural anthropology texts (e.g. Bonvillain, 2006; Lavenda and Schultz, 2013; Ferraro, 2006) reveals a similarly consistent set of themes explored by that discipline (Table 1, Column 2; note the subjects / rows in the columns do not correspond to one another, but simply indicate the sequences of topics as they are commonly presented in such texts). Broadly speaking, these are the topics that cultural anthropology is used to explain. They recur here (and through the history of academic anthropology) not because they are not just concerns of the present day but because human behavior is not random; rather it is to some variable degrees patterned in ways that address the essential requirements of biologically- and behaviorally-modern humanity. They are here identified because these texts' organization—just as the organization of topics in introductory mathematics or physics texts, for example—reveals the overarching issues explored.

**Table 1:** Stable Elements in Modern North American Evolutionary Biology and Cultural Anthropology Texts. Analogues are not implied between items adjacent in the two columns.

| Evolutionary Biology | Cultural Anthropology |
| --- | --- |
| Origins | The Culture Concept |
| Variation | Language |
| Adaptation | Kinship / Descent |
| Diversity | Power Relations |
| Heredity | Sex and Gender |
| Novelty | Equality and Inequality |
| Classification | Religion |
| Biogeography | Economy / Subsistence |
| Speciation | Myth, Ritual and Symbol |

More specifically, such patterning derives at least in part from what G.P. Murdock (1940)—at the mid-20th-century origins of modern anthropological theory—recognized as several universal aspects of human culture (e.g. culture (1) is learned, (2) is socially transmitted with symbols, (3) satisfies or attempts to satisfy basic needs, and (4) is adaptive) (pp. 364-368). Specifically, the facts that modern humans are large, highly-social, bipedal primates living in certain ecosystems, and use culture more so than biology to adapt, have conditioned the essential problems (or, we might say, shaped the selective environment) that culture must solve (e.g. social organization, rules of inheritance, etc.). This adaptive bent, however "…by no means commits one to an idea of progress, or to a theory of evolutionary stages of development, or to a rigid determinism of any sort. On the contrary…different cultural forms may represent adjustments to like problems, and similar cultural forms to different problems." (pp. 367). Despite some critiques of Murdock's claim of universals, Brown (2004) has provided evidence, compelling to the authors, of "A small number of causal processes or conditions [that] account for most if not all universals…(1) the diffusion of ancient, and generally very useful, cultural traits, (2) the cultural reflection of physical facts, and (3) the operation, structure and evolution of the human mind" (pp. 50). Table 2 informally identifies some of the more common domains of behavior guided by cultural information; these include some of Brown's own list of 'human universals' (e.g. body adornment, production and use of tools, metonymy [symbolism], age segregation and so on) and others proposed by other anthropologists.

**Table 2:** Some Cross-Culturally Observed Domains of Cultural Influence, or 'Human Universals'.

| Domain | Concept | Examples |
|---|---|---|
| **Language** | Specific spoken and gestural (bodily) systems of communication, including vocabularies and grammars. | Some languages assign gender to nouns, while others do not. |
| **Concepts of Space** | Concepts of distance; scales of interaction, from individual to community and extra-communal; also, units considered appropriate for measurement of space. | Some cultures reckon traveling distance in 'moons', e.g. nights of travel required to reach a destination, while others use more formal units such as leagues or kilometers. |
| **Concepts of Time** | Concepts concerning the passage of time, e.g. how it is reckoned with units considered appropriate. | Cyclical time is a fundamentally different concept than linear time; counting up from some distant event or down to some future event. |
| **Ethics** | Concepts of right and wrong, justice, and fairness. | Some cultures execute murderers, while others do not. |
| **Social Roles** | Rights and responsibilities differ by categories such as age (child, adult), gender (man, woman), and status (peasant, King). | Cultures differ in the ages at which people take on certain rights and responsibilities, and specifically what those rights and responsibilities are. |
| **The Supernatural** | Concepts regarding a universe considered fundamentally different from daily experience. | Different cultures worship different gods, goddesses, and other supernatural entities. |
| **Styles of Bodily Decoration** | Human identity is often communicated by bodily decoration, either directly on the body or with clothing. | Some cultures heavily tattoo the body while others communicate identity more with clothing styles. |
| **Family Structure** | Concepts of kinship or relations | Some cultures are polygynous, |
| | between kin, and associated ideas such as inheritance. | where males have several wives, and some are polyandrous, where females have several husbands. |
| **Sexual Behavior** | Regulation of sexual behavior, including incest rules. | Cultures differ in the age at which sexual activity is permitted. |
| **Food Preferences** | Concepts of what are appropriate food and drink in certain situations. | Some cultures eat certain animals while others consider them unfit to eat. |
| **Aesthetics** | Concepts of ideals, beauty, and their opposites. | Some cultures value visual arts more than song, and vice versa. |
| **Ultimate Sacred Postulates** | Central, unquestionable concepts about the nature of reality. | Some cultures consider all life to be a reincarnation of discrete beings in the past, while others envision human passage to entirely another domain after death. |

Table 2, then, may serve as a guide to stable or universal elements of culture that can profitably be investigated with an EES approach free of strict Darwinism or the Modern Synthesis, and rather informed by the richer theory of the EES. While there is no general consensus regarding what precisely constitutes the EES, it certainly includes multiple genomics-informed facets including developmental genetics, plasticity, phenotypic integration, niche construction, multilevel selection theory, mutualisms and regulatory evolution (Smith and Ruppell, 2011; Laland et al., 2015). These are outlined in Table 3 and Figure 1.

**Table 3:** Characteristics of Darwinian, Modern, and Extended Evolutionary Conceptions.

| Darwinian Evolution (1859-1950) | Modern Synthesis (1950-2000) | Extended Evolutionary Synthesis (2000-present) |
|---|---|---|
| Replication (heredity) | Mendelian inheritance | Inclusion of more modes of heritability; e.g. horizontal gene transfer, epigenetics, culture. |
| Gene-->Protein-->Phenotype | Developmental schedules. | Evolution of developmental regulation |
| Variation | Mutagenesis rare, by by 'zap' effectors e.g. cosmic rays | Mutagenesis common, as a result of mutation-repair failure. |
| Selection | Natural selection on individuals. | Mutualisms and symbioses; selection on multiple scales, niche construction involving 'self-selection'. |

---------------------------------------------------------------------------------------------------------------

**Figure 1:** Conceptions of Biological Variation and Change from Medieval to Extended Evolutionary Synthesis.

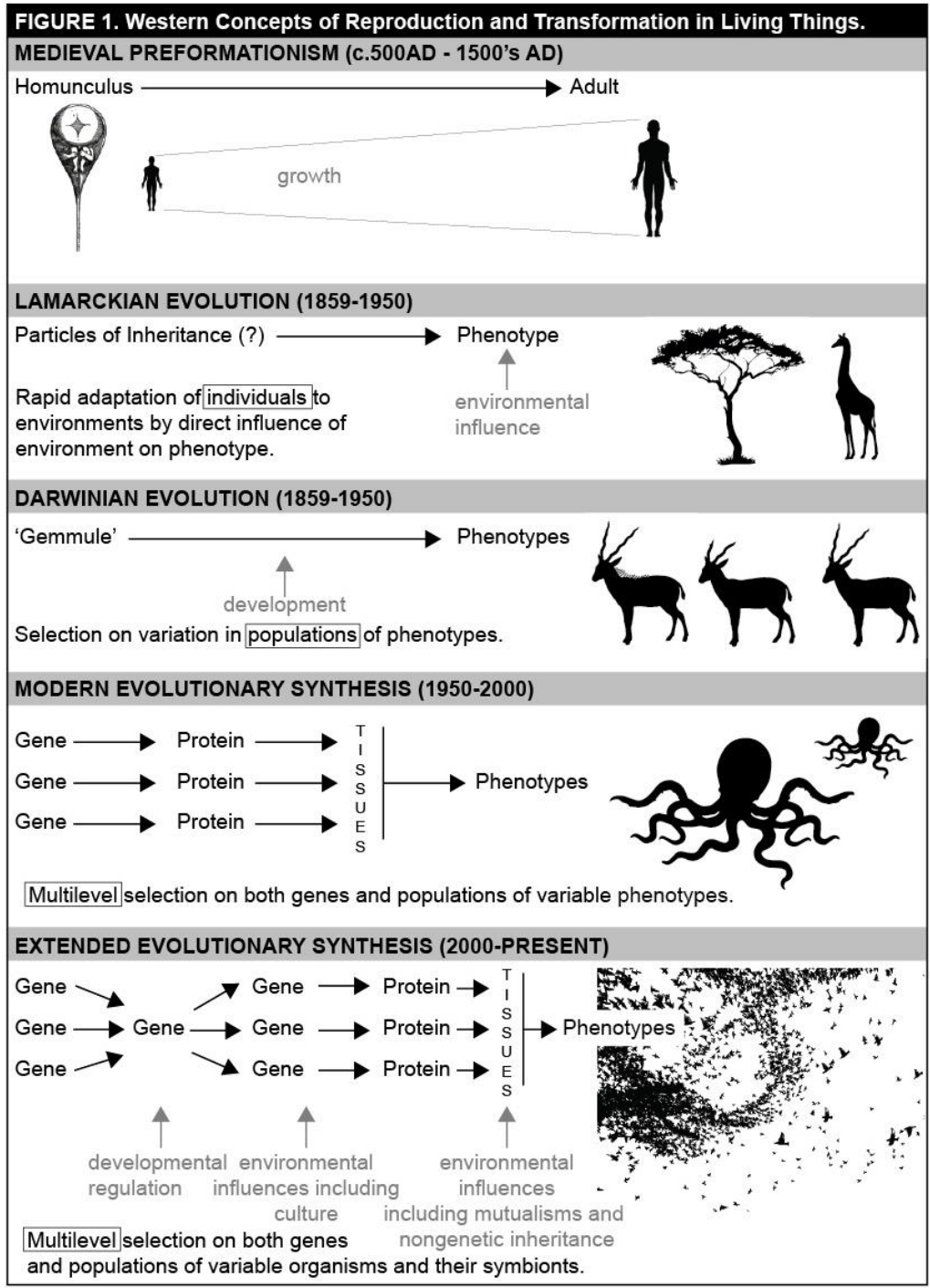

**3.     Heuristic Devices, Analogy and Metaphor**

At this point, it is important to address the serious danger attempting to explain the frequency of cultural traits in terms of a biologically-derived conception of fitness. We suggest that this issue may account for the sterility of the highly reductionist approaches to cultural evolution—including evolutionary psychology and memetics—of the 1990-2000 era (e.g. Cosmides and Tooby, 1997). In biology, characteristics that confer lower fitness can persist for several reasons, i.e., they may be 'hitchhiker genes' that piggyback alongside other genes that confer adaptive benefit (Smith and Haigh, 1974). Similarly, an 'optimizing' approach to cultural evolution is problematic because maladaptive cultural traits can hitchhike alongside beneficial ones (Gabora, 1997), and even persist for centuries in certain conditions (Edgerton, 1991). This persistence may reflect that a given cultural trait may be adaptive for some, and not for others (e.g., slavery, see Donald, 1997 and Wolf, 1982) or adaptive in some contexts, and not others (Pierce and Ollason, 1987).

It is easy to misleadingly overextend analogies between cultural and biological processes (Claidière and André, 2011; Mesoudi, 2015). Nevertheless, cultural evolution is not simply 'like' biological evolution; it *is* an evolutionary process. Computational models of cultural evolution exhibit not just the cumulative, adaptive, open-ended change that defines an evolutionary process, but other key attributes such as epistasis, drift, overdominance, and underdominance, as well as incorporating phenomena unique to culture, such as the capacity to learn trends and use them to bias the generation of novelty, and the capacity to mentally simulate outcomes without having to actualize or manifest them (Gabora, 1995; 2008). Where it is not (or where the issue is for the moment unclear), we may still make use of metaphor in discussing biological and cultural evolutionary processes; useful metaphors can stimulate exploration, communicate essences, serve as aids to memory and aid in experimental design, but we must beware not to reify them, or develop a false sense of understanding when using them; for Hoffman (1980), "…scientists…can be quite well aware of the differences between concepts given by the theories and concepts suggested by metaphors [which are] used to explore nature and to lead to modification of principles" (pp. 403). We know that musical notes are not played by DNA code, but a useful metaphor for the revelations of developmental genetics is that is that the genome is more 'like a jazz score than a blueprint' (Porta, 2003), an heuristic device that does not build a "just so" story (see also Boone and Smith, 1998 and Smith et al., 2001). As Sober (2006) points out, "descriptors singled out for treatment in science always abstract from complexities. If there is an objection to the descriptors used in models of cultural evolution, it must concern the details of how these models are constructed, not the mere fact that they impose a descriptive framework of some sort or other" (pp. 487).

We believe the EES may indeed apply the understanding of cultural change in that, when appropriately conceived (as explored above), cultural evolution is indeed an instance of general evolution, in the full, nontrivial sense of cumulative, adaptive, open-ended change.

## 4.     Four Domains of the Extended Evolutionary Synthesis and How They Facilitate Evolutionary Models of Culture

Below we introduce four domains of the Extended Evolutionary Synthesis in which there are clear and significant implications for studies of cultural evolution.

*4.1     Horizontal Gene Transfer and the Lamarckian Dimension of Cultural Evolution*

Cultural information may obviously be transmitted horizontally (among a generation) as well as vertically (between generations), allowing individuals to adapt more quickly to changing selective pressures than is possible under either a strictly genetic mode of transmission or system that includes only individual trial-and-error learning (Alvard, 2003, but see also Mesoudi et al., 2004). Anthropologically, rapid results of horizontal cultural information transfer are invoked in the term *ethnogenesis* (Tehrani and Collard, 2002; 2009; Collard et al., 2006), which stands in contrast to phylogenetic evolution of biological information (Mace and Holden, 2005).

## 4.2    Mutualisms: Gene-Culture Coevolution

Organisms in a mutualistic pairing can hitchhike with one-another even when only one possess a gene that is under positive selection (Nadell and Foster, 2012). In the same way culture and genetic information systems co-evolve, as noted since the 1990's (Durham, 1982; Feldman and Laland, 1996). Synthesizing work continues in this field (Richerson and Boyd, 2005), particularly as the human genome is understood in finer, functional detail, as in the well-studied case of lactose tolerance in Northern Europeans in which genes coevolved with cultural (dietary) norms (Beja-Pereira et al., 2003; Laland et al., 2010). Social mutualism extends to increased sociality and importance of smooth social navigation among higher primate groups and *Homo* in particular (Tomasello et al., 2005) as pointed out generally in Dunbar's social grooming hypothesis (Dunbar, 1991), featured in a transition in hominin evolution from social close-*kin* selection to close-*group* selection (Foley and Gamble, 2009). Insights from the world of biological mutualisms should help in explicating and explaining the coevolution of mosaic traits in early hominin evolution, such as the hand morphology / tool use / enculturation suite (Marzke, 2013; Hünemeier et al., 2012).

Recent work in this domain include investigations of gene-culture interaction on the rate of evolution (Hünemeier et al., 2012), the role of gene-culture interactions in geographically-restricted adaptation over the last 50,000 years (Laland et al., 2010), gene-behavior coevolution in the case of the origins of language (Aoki, 2001), the evolution of social norms (Gintis, 2003) and the global, early-Holocene experiments with plant and animal domestication; in a recent work on the European Neolithic (Zeder, 2008) it has been noted that multiple taxa were significantly coevolving. Foundation work has yet to be done, however, and some call for refining our definitions and exploration of the relationships of organism and its environment, and recent work explores the distinction between *idea-centered* and *organism-centered* cultural evolution (DeBlock and Ramsey, 2015). Some of these issues are also informed by *niche construction theory* (Scott-Phillips et al., 2013; Odling-Smee et al., 1996) that has been explicitly applied to cultural adaptation. For example, Scott-Philips et al. (2013) illustrate the significance of viewing the domestication of dairying animals not as simply a "background condition" to human genome evolution but a proactive human action, a manifestation of a human "propensity to bias selection pressures'' resulting in allele frequency change. Niche construction theory, in this way, provides an updated vocabulary and perspective on evolution particularly suited to human evolution, which has been uniquely proactive, at least since the origins of behavioral modernity ca. 100,000 years ago.

## 4.3    Gene Regulation and Expression and the Regulation of Cultural Behavior

Developmental evolutionary studies have a long history in biology (Laubichler and Maienschein, 2008) and currently there is much focus on genes regulation of other genes, according to a schedule or on response to environmental signals (in human evolution, significantly altering the expression of functionally conserved proteins and regulatory gene mutation; see Capra et al., 2013; Carroll, 2008), as in the cases of epigenetic factors associated with obesity, cancer, cardiovascular disease, type 2 diabetes and colon cancer (Shen et al., 2007).

Similarly, in the cultural information system, information flow is regulated by biological and cultural 'valves': the ability to filter cultural signals may be mediated by complex psychological developments, such as the becoming sensitive to approval or disapproval and to outwardly approve or disapprove of others, this disposition becoming the 'regulatory switch' allowing or prohibiting cultural expression (see section 5, below); cultural developmental schedules, e.g. rites of passage, also regulate cultural expression (Greenfield et al., 2003). Acerbi et al. (2014) have developed models of cultural evolution in which cultural "regulators" allow for innovation (see section 4.4, below) to be modelled, and they explicitly introduce cultural behaviors analogous to regulator genes. Developmental schedules regulating the expression of cultural information might well be investigated in the phenomena of rites of passage, cognitive development stages, language acquisition and lifetime-scale enculturation processes.

## 4.4    Mutagenesis, Phenotypic Variation and Cultural Innovation

The field of mutagenesis is currently on its head; while mutation was once considered a rare and specific result of such limited variables as cosmic ray bombardment and mechanical deformation of the DNA, it is now seen as a continual process with many authors, and in fact largely the result of *DNA repair mechanism failure* (Friedberg, 2006). However one defines evolutionary novelty (discussed in Brigandt and Love, 2010) it is the origin of variation in evolutionary information, biological and cultural (e.g. Bender and Beller, 2014). In both systems of evolutionary change, innovation is variation from established patterns, which include (in biology) conserved genes (Woolfe et al., 2005) such as those in the homeobox clusters and in human cultures might first be examined by cross-cultural study of 'human universals' (see Table 2); another promising starting point is the study of highly-conserved words (e.g. Pagel et al., 2013) and genuine cultural universals (e.g. Smith, 2011).

Just as variation and population size are important biologically, in culture the interplay between demography, cultural innovation, and fitness is significant; in simulations, small populations are more likely to retain less beneficial cultural innovations producing low equilibrium fitness (Shennan, 2001). In larger populations, sampling effects (in terms of both lateral and vertical transmission of fitness lowering innovation) are lessened, giving those populations a selective advantage; these simulations also suggested a positive causal correlation between population size and fitness values associated with innovation.

Early on, Barnett (1953) examined cultural innovation, placing it at the center of cultural evolution. Recent simulations suggest that strategies where agents adopt conservative, culture reproducing actions mixed with individual innovation, depending on environmental settings, greatly increase overall fitness (Castro and Toro, 2014; Wakano and Miura, 2014). While still relatively simplistic, such models have the potential to more lead to more nuanced discussions of agency, innovation, and cultural stability. 'Cultural backgrounds' act as constraints on innovation (Rueffler et al., 2006; Bryson, 2014; Burns and Dietz, 1992) much as the *bauplane* (*sensu* Gould and Lewontin, 1979) sets biological constraints to biological variation.

## 5.       Summary

Bamforth (2002) summarized the potentials and perils of applying evolutionary theory to cultural studies. Although he was speaking specifically of strictly interpreted Darwinian theory in archaeology, he warned against uncritically applying terminology. The new 'pluralistic' model of heredity is still relatively young. Many of the mechanisms involved with non-genetic inheritance are not yet fully understood (Bonduriansky, 2012; see also the new *Journal of Non-Genetic Inheritance*], nor is the path to integrate the various emerging biological explanations into a cohesive whole apparent (Day and Bonduriansky, 2011). It is one of the ironies in the history of anthropology that even though many of the process involved with cultural evolution are perfect examples of the concepts research in biological heredity are trying to understand now, attempts to conform to the Modern Synthesis model of evolution have kept anthropology from taking the leading role of studying cultural evolution that it might otherwise have taken (Tomczyk, 2006). Methods originally designed for the study of quantitative genetics can be applied to the study of culture. Creating models to test hypotheses of cultural evolution is of utmost importance. Many authors advocate the adoption of neutral models as baselines for null hypothesis testing (e.g. Bentley et al., 2004; Bentley et al., 2007; Lipo et al., 1997; Crow and Kimura, 1970; Zhang and Gong, 2013; Vogt, 2009). Mathematical models must, in addition, have explicitly stated assumptions and clearly defined, realistic estimated parameters (Bell and Spector, 2011; Rakyan et al., 2002). Similarly, in culture similar studies could be made using groups that have recently split from a common origin. A recent overview of human behavioral ecology noted that evolutionary studies should incorporate the mechanisms, development, phylogeny, as well as function, although until recently not enough focus has gone toward mechanism (Borgerhoff Mulder and Schacht, 2012). The thoughtful exploration of the pluralistic model of heredity as it applies to culture can lead to new avenues to explore culture, just as a biology can benefit from the unifying of study of both genetic and nongenetic inheritance (Day and Bonduriansky, 2011). Much as the ongoing debate in biological evolution over "true" importance of Darwin's legacy continues (Ingold, 2007), debates about the "right" evolutionary model for culture (if one exists), are far from settled. The debates between proponents of various anthropological theories (let alone the distrust of those who add to the discussion from outside the field [Tomczyk, 2006]) hamper insight into what is important about cultural evolution much as methodological differences have hampered insight into human origins.

At the beginning of this paper we referenced Claidière and Andrè (2011) who question the notion of applying population models of transmission to culture. They end with a call for the inclusion of "novel concepts and mechanisms" in the analysis of cultural evolution; we feel the EES is supplying such concepts and mechanisms as illustrated in this paper and ongoing research (e.g., see Andersson and Read, 2016; Gabora, 2013; Smaldino and Richerson, 2013; Sterelny, 2016).

Overall, the EES provides evolutionary models of culture an alliance with and legitimate access to a century or more of genuinely evolutionary studies; the models, debates, larger and smaller confirmations and disconfirmations of biological evolutionary studies may now be accessed and evaluated for their applicability to the world of cultural evolution studies. This is what the EES supplies; it also rescues evolutionary approaches to culture: the earliest, Social-Darwinian approach was rightly rejected as unilineal and over-deterministic, but unfortunately this clouded efforts to build more progressive models in the last five decades (see

*Appendix 1* for historical reviews). This cloud may be lifted by application of the EES as described here.

## Acknowledgements

We thank the many authors of the Extended Evolutionary Synthesis for their long and largely-unreported but important work in bettering our understanding of the human species. This research was supported in part by a grant (62R06523) from the Natural Sciences and Engineering Research Council of Canada to Dr. Gabora.

## Appendix 1

While the history of evolutionary approaches to cultural change is a separate topic, readers may find reviews in Mesoudi, Whiten and Laland (2006), Steele, Jordan and Cochrane (2010) and Creanza, Kolodny and Feldman (2017). Direct explorations of the topic (following the Modern Evolutionary Synthesis include), including both anthropological and nonanthropological theoretical backgrounds, include Cavalli-Sforza and Feldman (1981), Hull (1988), Basila (1988), a very ambitious and thorough treatment in Durham (1991), Aldrich (1999), Fog (1999), Kluver (2002) and Blute (2010) and a recent issue of the *Proceedings of the National Academy of Sciences of the United States,* "The Extension of Biology through Culture" (Whiten et al, 2017).